\documentclass[a4paper,twoside,11pt]{article}
\usepackage{a4}
\usepackage{graphics}
\usepackage{epsfig}
\def\utw{\smash{\rlap{\lower5pt\hbox{$\sim$}}}}
\def\udtw{\smash{\rlap{\lower6pt\hbox{$\approx$}}}}

\def\AJ{{\it Astron. J.} }
\def\ARAA{{\it Annual Rev. of Astron. \& Astrophys.} }
\def\ARNPS{{\it Annual Reviews of Nuclear \& Particle Science} }
\def\ApJ{{\it Astrophys. J.} }
\def\ApJL{{\it Astrophys. J. Letters} }

\def\AA{{\it Astron. \& Astroph.} }

\def\AAL{{\it Astron. \& Astroph. Letters} }

\def\Nature{{\it Nature} }
\def\MNRAS{{\it Month. Not. Roy. Astr. Soc.} }

\def\etal{{\it et al.}}

\begin{document}

\begin{center}
\Large{\bf Single and binary Black Holes and their active environment}

{\it Peter L. Biermann$^{1,2,3}$, Mihaela Chirvasa$^{1,4}$, Heino
Falcke$^{1,5}$, Sera Markoff$^{1,6}$, and Christian Zier$^1$}

{$^1$  Max-Planck-Institute for Radioastronomy, Auf dem
H{\"u}gel 69, D-53121 Bonn, Germany\\
$^2$  Department for Physics and Astronomy, University of Bonn, Germany\\
$^3$  Laboratoire de Physique Th{\`e}orique et Hautes Energies, \\
Universit\'e Pierre et Marie Curie (Paris VI) et Denis Diderot (Paris
VII), 4, Place Jussieu, F-75252 Paris, France\\
$^4$  Max-Planck-Institute for Gravitational Physics, Am
M{\"u}hlenberg 1, D-14476 Golm near Potsdam, Germany\\
$^5$  Physics Department, University of Nijmegen, Nijmegen, The
Netherlands\\
$^6$  Center for Space Research, Massachusetts Institute for Technology,
77 Massachusetts Ave., Cambridge, MA  02139, USA}
\end{center}

%                                                            
%      draft PeBi, version of Nov 7, 2002, plbiermann@mpifr-bonn.mpg.de
% ... still with ps-errors and printing problems with Fig 9+10 ....
%      sera@space.mit.edu,albrecht.dress@arcor.de
%      tschweiz@ifae.es,mihchi@aei-potsdam.mpg.de
%      hfalcke@mpifr-bonn.mpg.de,chzier@mpifr-bonn.mpg.de
%      
%

\section{Abstract}

In this short review we describe some of the latest endeavours to
understand the activity around Black Holes.  Black Holes have now
been observed on many mass scales, from about 5 solar masses all the
way to almost $10^{10}$ solar masses.  Stellar size Black Holes
appear to exist all through galaxies, while more massive Black Holes
appear to occur only in the centers of galaxies.  First we outline some
efforts to understand accretion disks and especially jets around
stellar-size Black Holes, also known as microquasars.  Here it has been
possible to demonstrate that a large part of the electromagnetic emission
observed can be interpreted as arising from the jet; this explains at
once all spectral features and their variability.  Second we dwell on the
concept that merging galaxies naturally lead to merging Black Holes. 
Here we emphasize two aspects:  a) the torque exerted by the binary
Black Holes carves a torus like distribution out of the stellar
population near to the Black Hole binary; the sub-population of red and
blue super giants with stellar winds may have their winds turned into
tails, which as an ensemble may make up the ubiquituous torus, inferred
from X-ray, infrared and optical polarization observations.  b)  We
consider the last stages of the Black Hole binary merger, taking into
account the angle between the spin of the primary Black Hole, and the
orbital spin of the second Black Hole.  We show that the loss of orbital
angular momentum is very strongly spin-dependent; for large angles
between the two spins the angular momentum loss is strongly inhibited,
allowing for spin flips of the primary Black Hole, as it merges to become
the new combined Black Hole.  This spin flip preserves a high angular
momentum relative to the maximum allowed.  This ensures that both before
and after the merger the accretion disk may reach to very small distances
from the central Black Hole, with very high local temperatures right near
the base of the jet:  This is especially interesting in the case that
forming the jet requires the formation of an ADAF like ring near the
inner edge of the disk, as suggested by some earlier work.  It also may
have consequences for the initial hadronic interactions right near the
base of the jet.  Finally, this may also have important implications for
the discovery of gravitational radiation bursts from the merger of black
holes; the spin dependence needs to be taken into account.

\section{Introduction}

Black Holes have now been found on all mass scales, from stellar size to
about $10^{10}$ solar masses.  They are very common, with the more
massive Black Holes all apparently occurring in the center of galaxies;
almost all galaxies seem to harbour a central Black Hole.  When they
accrete, they display an enormous variety of phenomena, accretion disks,
relativistic jets, broad and narrow emission line clouds, and tori of cold
absorbing material.  Since the stellar size Black Holes, in a mass range
between about 5 and 10 solar masses, occur rather closer to us than any
other more massive Black Hole with large accretion power, they can be
used to study the underlying physics.  So we discuss in the first part of
this review some recent work, where we tested the hypothesis that the
underlying physics may be similar, if not the same, on all mass scales of
Black Holes.  We find that stellar size Black Holes and their activity
are well described by a scaled down version of the concept developed for
more massive Black Holes.  Then we move on to more massive Black Holes,
and note that mergers of galaxies naturally lead to binary Black Holes.  
Binary Black Holes eventually merge, and when they merge, their spin
becomes critically interesting for the last stages of their
gravitational wave emission.  In order not to explode the list of
references, we have been very parochial and just mention our own
work, and a few pertinent reviews, and some limited set of references.

\section{Microquasars, Active Galactic Nuclei and High Energy
Particles}

In a series of papers Falcke et al. have developed a jet-disk-symbiosis
picture, \cite{jd-1,jd-2,jd-3}:  In these papers it was possible to show
that with just mass and energy conservation and a few simple scaling
arguments the emission from the jet and the disk could successfully be
modeled; it was shown, e.g., that equipartition in the comoving frame
turned out to be a good approximation, yielding the most efficient jet
system.  In the last three years we have been able to show that the
jet-disk symbiosis picture developed is also applicable to the binary
star Black Hole systems.  These are stellar binary systems, where one
partner is a normal star, and the other one a Black Hole.   These systems
are also known to exhibit an accretion disk, fed by mass transfer from the
partner star, a radio jet, which is also relativistic, and therefore they
have been dubbed ``microquasars", \cite{MiRo00}.  There has been a lot of
activity centered on such systems, e.g., 
\cite{MiRo98,MDC98,MiRo99,MiRo00,DMR00,MDMRG01,KBRM02,RKBM02}.

The recent observations of microquasars with the X-ray satellite CHANDRA
provided a welcome opportunity to test the physical concepts of
our jet-disk symbiosis concept originally developed for more massive
systems.  In the last three years we have been able to show that almost
the entire electromagnetic spectrum can be described by emission from the
jet, and only a small part of the observed emission arises from the
accretion disk, and possibly from a small disk corona.  The variability
and the spectrum could be matched with a simple description of
synchrotron and inverse Compton emission from the jet, using shock
acceleration as one key concept, \cite{KFM02,MFF01,MNCFF02}.  The model
then indicates the location of the shock and its physical parameters, as
well as the phase space distribution of the energetic electrons and/or
positrons.  Fig. 1 shows one example for our spectral fits for a
microquasar.

%%%%%%%%%%%%%%%to insert figure 1 %%%%%%%%
\begin{figure}
\centering\rotatebox{0}{\resizebox{14cm}{!}%
{\includegraphics{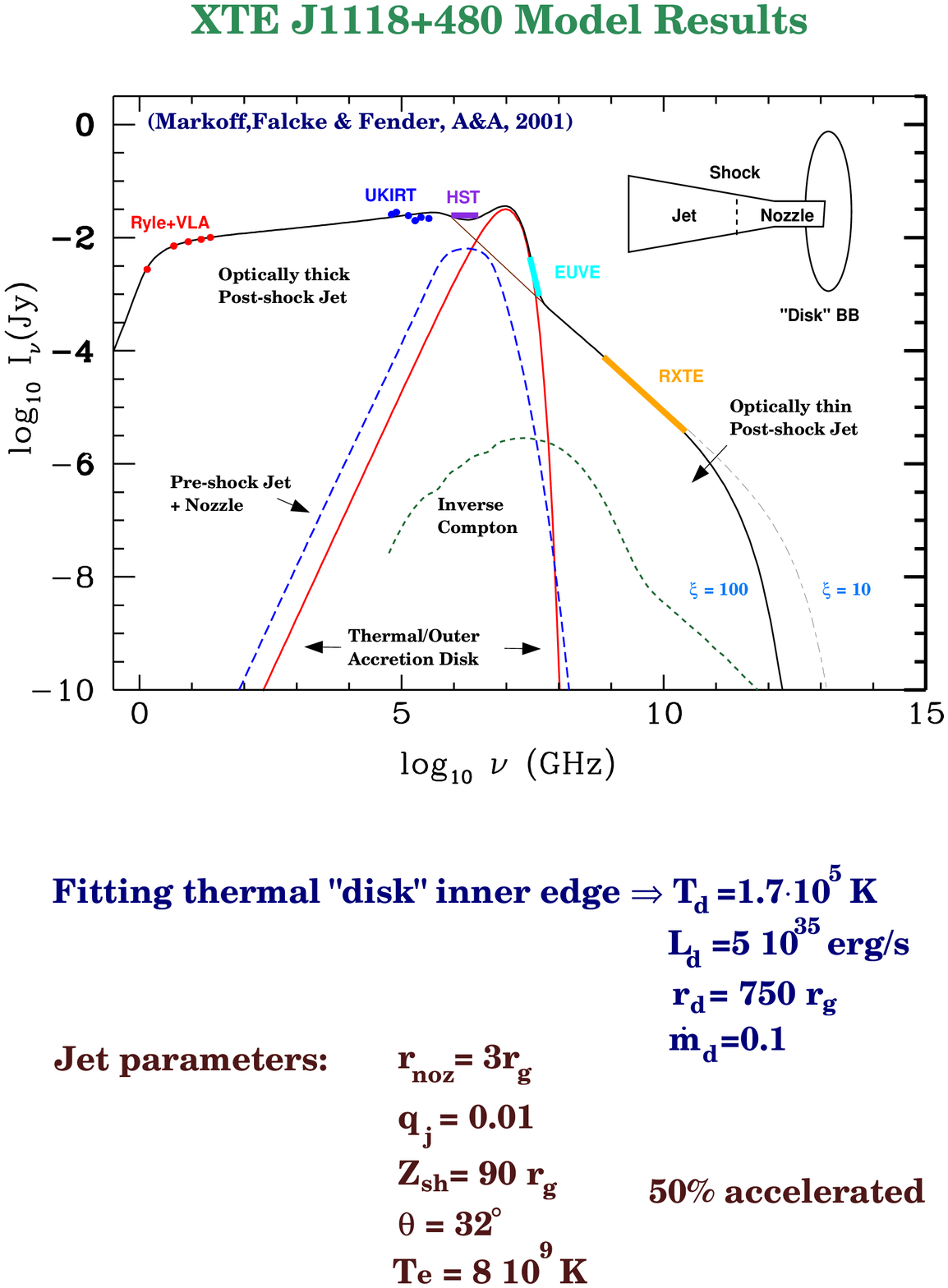}}}
\caption{Our jet-disk symbiosis model fitted to the CHANDRA data, using
the microquasar XTEJ1118+480 as an example; we also show the geometry of
the model.} 
\end{figure}
%%%%%%%%%%%%%%%%%%%%%%%%%%%%%%%%%%%%%%%%%

These successful and conceptually simple fits have now been extended
to low luminosity Active Galactic Nuclei (AGN),
\cite{FaMa00,MFYB01,YMF02,YMFB02}, and provide now a backbone for 
studying the hadronic interactions in jet-disk systems of all powers.
Microquasars are now a very well-studied testbed for all processes
expected to take place in very powerful sources, including high energy
hadronic and leptonic processes.

\subsection{Gamma Ray Bursts}

Similarly, Gamma Ray Bursts provide one further test for Black Hole
formation in stellar binary systems, and also for cosmological evolution,
as we have shown and continue to investigate. One possibility to  explain
Gamma Ray Bursts is to adopt the existing binary systems such as SS433 as
predecessors, a massive early type star in a binary system with a neutron
star, which is surrounded by an accretion disk and produces a jet,
\cite{GRB-1}.  In  such a system all parameters are well known, there is
little freedom as to the magnetic field strength and other key parameters
of the system.  When the neutron star is pushed over the brink of
instability and becomes a Black Hole, a gigantic explosion pushes an
ultrarelativistic shock wave along the pre-existing channel, the jet.  It
has been possible to show that the emission from this shock can explain
all the gross features of Gamma Ray Bursts afterglows.  Furthermore,
connecting the star formation rate with the Gamma Ray Burst creation rate
leads to a cosmological evolution as a function of redshift, which is
sharply defined:  It rises steeply to a redshift near 1.7, and declines
only gently to higher redshift, in full agreement with the arguments from
other wavelengths.  Using thus the absolute scaling provided by this
quantitative fit to the numbers - taking selection effects into account,
following Petrosian, \cite{PeLe96} - leads then to a determination of the
possible contribution of Gamma Ray Bursts to the flux of cosmic rays,
\cite{GRB-2}:  Both from Galactic sources as from extragalactic sources
it is hard to see how their contribution could be significant.

\subsection{Radio Galaxies:  Ultra High Energy Cosmic Rays}

This work leads to the possibility to verify again our concept that radio
galaxies and their cousins, the relativistic jets pointed at us such as
BL Lac type sources and GHz peaked sources, provide the highest energy
particles known to us in the Universe,
\cite{BiSt87,EWNB98,Paris00,Erice00,ESA-CERN01,CR-APE02}.  All these
emission processes are likely to involve hadronic interactions and will
then be testable with very high energy neutrino observations,
\cite{LeMa00}.  Fig. 2 shows the example of the Galactic center low
luminosity AGN Sgr A*, see, e.g., \cite{MeFa01}.  Using the scaling of
magnetic field strength with other signs of power leads to an estimate of
the maximum energy of a particle that can be contained in the jet, along
its length or in its head, the hot spot - if there is one.  This suggests
that about
$3 \, 10^{21}$ eV is the maximum; allowing for some relativistic boosting
this limit may extend to
$10^{22}$ eV on the outside.

%%%%%%%%%%%%%%%to insert figure 2 %%%%%%%%
\begin{figure}
\centering\rotatebox{0}{\resizebox{16cm}{!}%
{\includegraphics{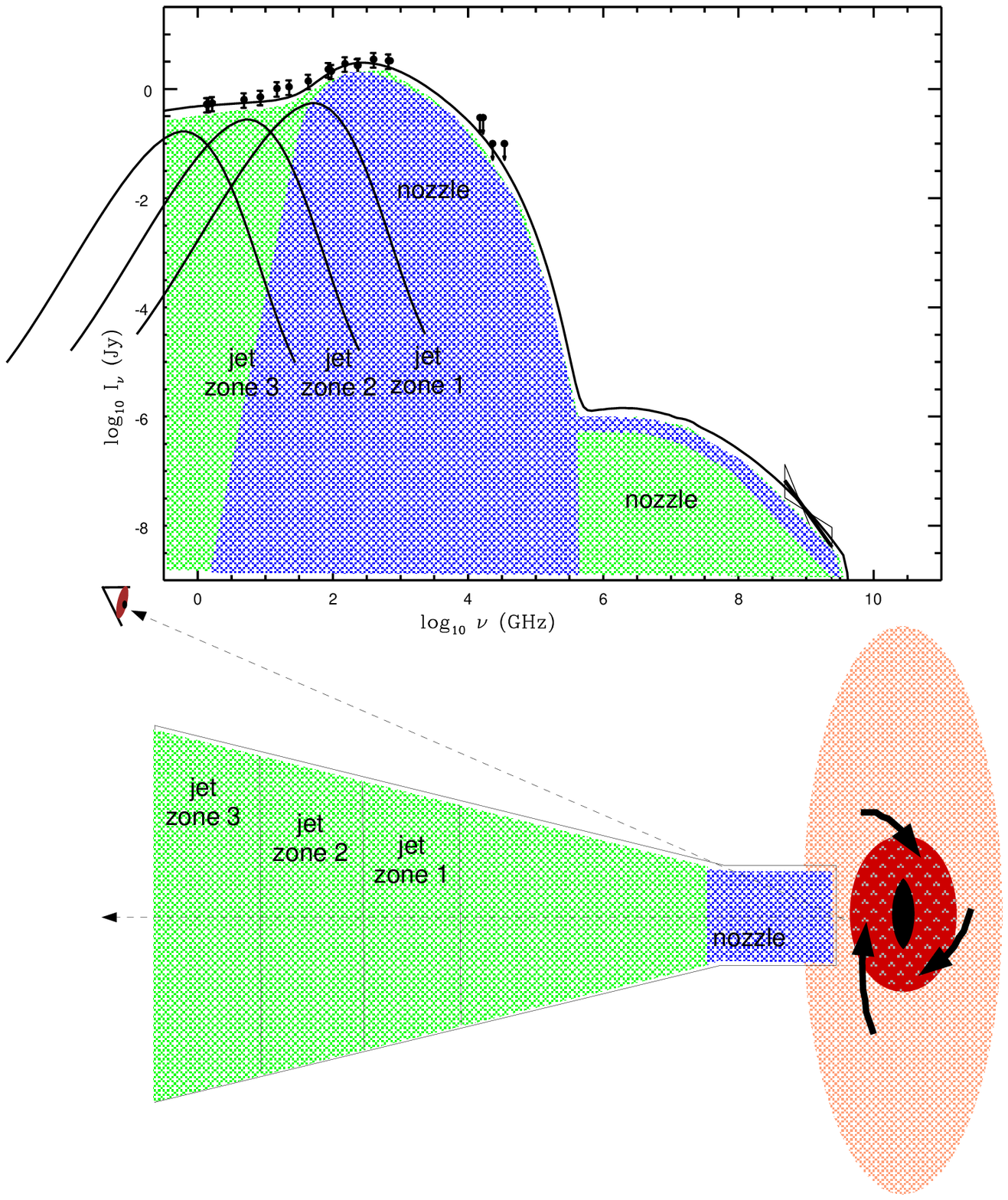}}}
\caption{Our jet-disk symbiosis model fitted to the CHANDRA data, using
the low luminosity Active Galactic Nucleus in our Galaxy Sgr A* as an
example; again we also show the geometry of the model.} 
\end{figure}
%%%%%%%%%%%%%%%%%%%%%%%%%%%%%%%%%%%%%%%%%

\subsection{Scaling laws for jet-disk systems}

What we learn then from applying the jet-disk symbiosis picture to
microquasars, low luminosity AGN, and also high luminosity AGN is that
the scaling laws are indeed a very good first approximation to the
complex physics undoubtedly ruling the physics of the activity around
single Black Holes.  This will present a welcome basis to study further
physical processes in relativistic jets, such as the acceleration of
protons and other charged nuclei, possibly yielding the highest energy
cosmic rays observed.

\section{Binary Black Holes}

Today the work by the Hubble Space Telescope has shown convincing
evidence that almost all galaxies have a central massive Black Hole,
e.g., \cite{KoRi95,FTA97}.  The mass of this Black Hole correlates well
with the total baryonic mass of the old spheroidal stellar population, and
correlates extremely well with the central velocity dispersion of the
central stellar population, \cite{FeMe00,MeFe01}.  In earlier work we have
shown that such correlations can be readily interpreted as the result of
the competition of star formation and accretion as a function of radial
distance in the disk of a galaxy, when considered as a gigantic accretion
disk, and subsequent mergers between galaxies,
\cite{WaBi98,WaBi00,WBW00,DSB00}.  As soon as two galaxies merge which
both have central Black Holes, the two Black Holes spiral in towards each
other under the rather strong influence of dynamical friction, with time
scales of order  $10^7$ to $10^8$ years.  This inspiraling slows down
considerably when the two Black Holes get as close as the core radius of
the central star distribution, of order a parsec or so. Again, this a
topic of wide interest, e.g., \cite{MMRB02,Me02,MeEk02}.

\subsection{The Central Torus}

From the early X-ray observations, \cite{LaEl82,Mu82}, the polarization
of the broad emission lines, \cite{AnMi85}, and later the far infrared
and submm/mm observations, it had been inferred that many, if not most
AGN harbour a central torus of cool obscuring material,
\cite{CKB89,CBKG89,SaMi96}; see fig. 3 for a sketch of the unified
scheme to understand AGN.  There are already a
variety of models for this torus.  First of all, one idea has been to
identify this torus with the irregular ensemble of merger remnants of
interstellar clouds,
\cite{SPN89}; another was to use radiation pressure to hold up the torus
material against collapse,
\cite{PiKr92,PiKr93}.  Also, a magnetic field model has been shown to be
viable \cite{LRB98}.  Here we show that a torus can be understood also as
the sum of many stellar tails, drawn out from an ensemble of red and blue
giant stars in a torus like geometry, \cite{Zi00,BBH-1}.

%%%%%%%%%%%%%%%to insert figure 3 %%%%%%%%
\begin{figure}
\centering\rotatebox{0}{\resizebox{14cm}{!}%
{\includegraphics{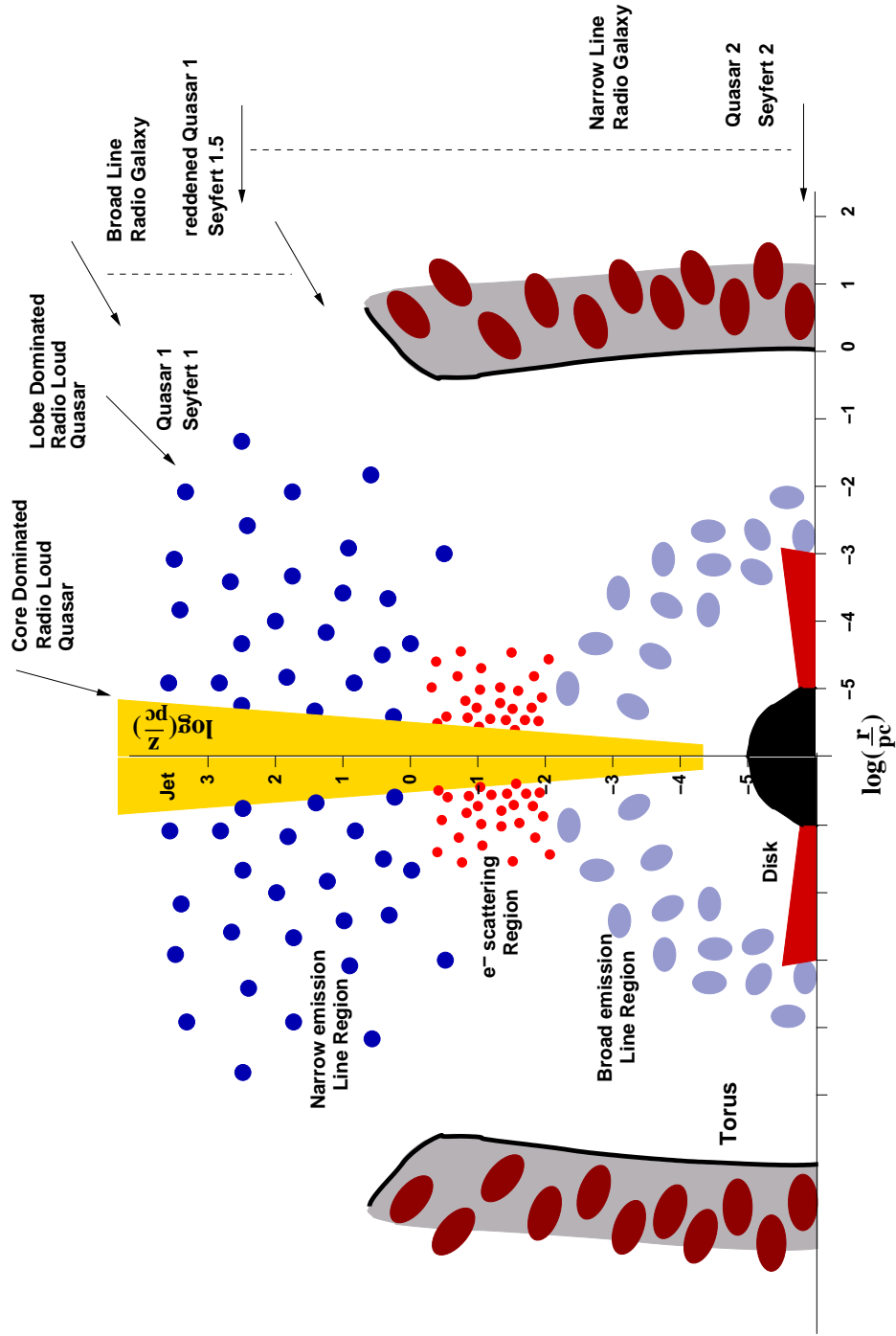}}}
\caption{The basic sketch of a unified model to understand Active
Galactic Nuclei, in a single Black Hole case.  The scales are
logarithmic, and the geometry is not fully mathematically
self-consistent in order to help the eye.  A spherical broad line cloud
distribution should really look more like a square with smooth corners in
such a graph, while the conical jet should be even closer to a
rectangular triangle in such a graph, with the 90 degree corner at the
origin.} 
\end{figure}
%%%%%%%%%%%%%%%%%%%%%%%%%%%%%%%%%%%%%%%%%

%%%%%%%%%%%%%%%to insert figure 4 %%%%%%%%
\begin{figure}
\centering\rotatebox{0}{\resizebox{14cm}{!}%
{\includegraphics{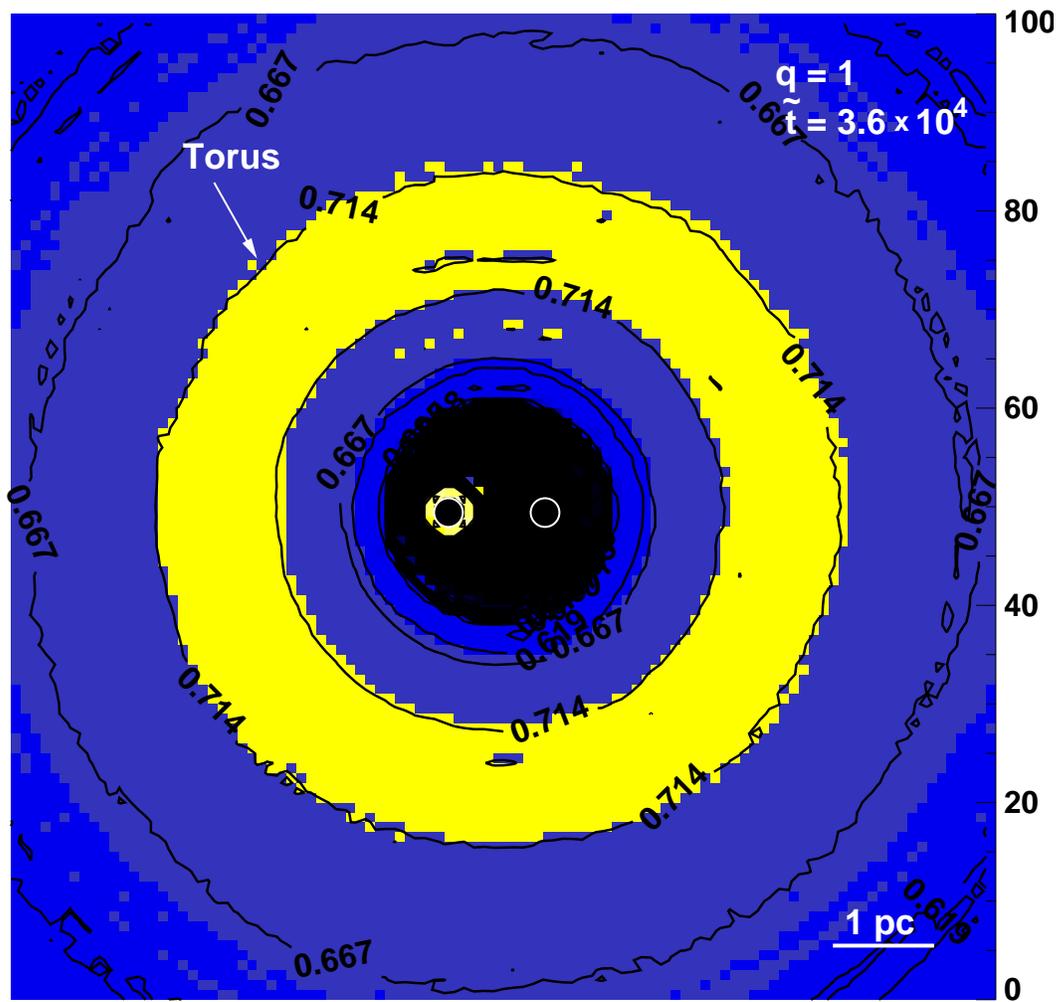}}}
\caption{The distribution of inner stars in the plane of the two Black
Holes, showing the broad belt in its own plane resulting from the torque
of the two Black Holes.} 
\end{figure}
%%%%%%%%%%%%%%%%%%%%%%%%%%%%%%%%%%%%%%%%%

%%%%%%%%%%%%%%%to insert figure 5 %%%%%%%%
\begin{figure}
\centering\rotatebox{0}{\resizebox{14cm}{!}%
{\includegraphics{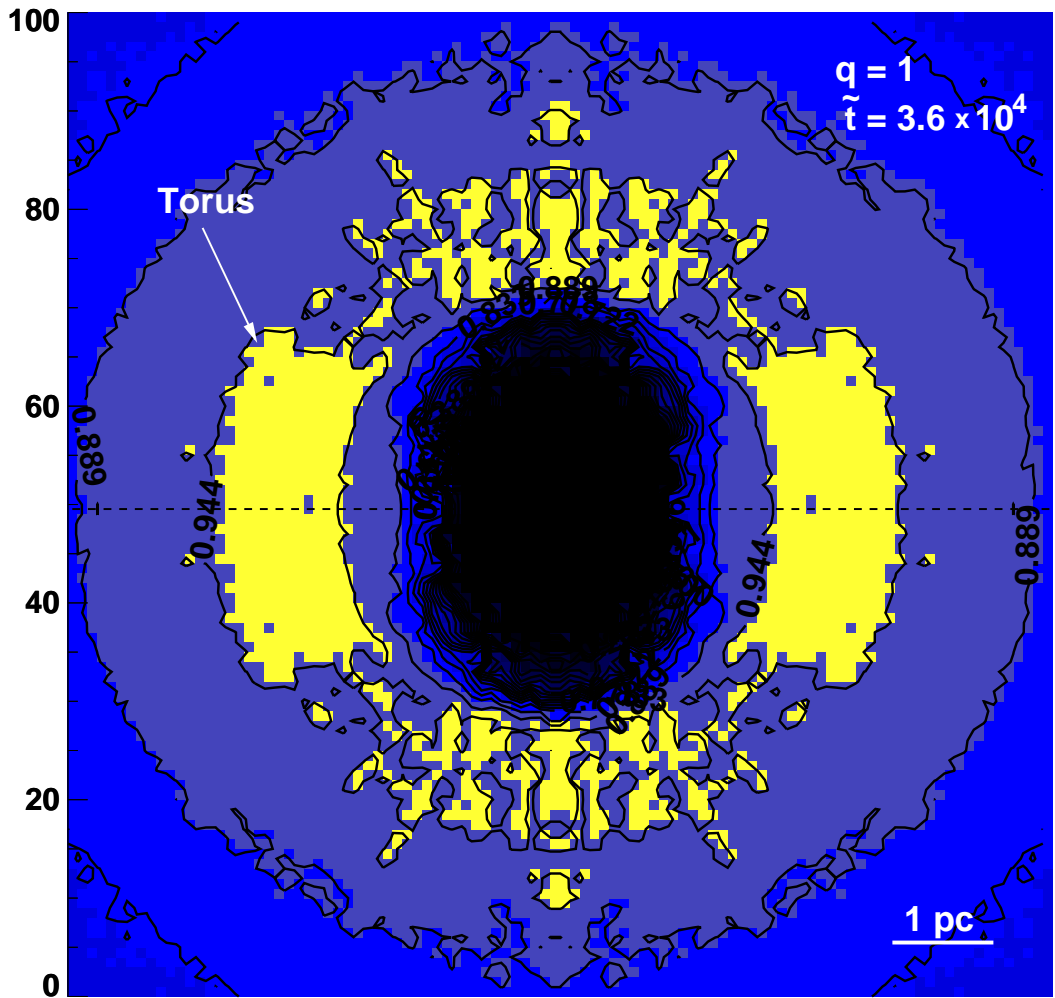}}}
\caption{The distribution of inner stars in a plane containing the
rotational axis of the orbit of the two Black Holes, showing the broad
belt in a cross-section resulting from the torque of the two Black
Holes.} 
\end{figure}
%%%%%%%%%%%%%%%%%%%%%%%%%%%%%%%%%%%%%%%%%

%%%%%%%%%%%%%%%to insert figure 6 %%%%%%%%
\begin{figure}
\centering\rotatebox{0}{\resizebox{14cm}{!}%
{\includegraphics{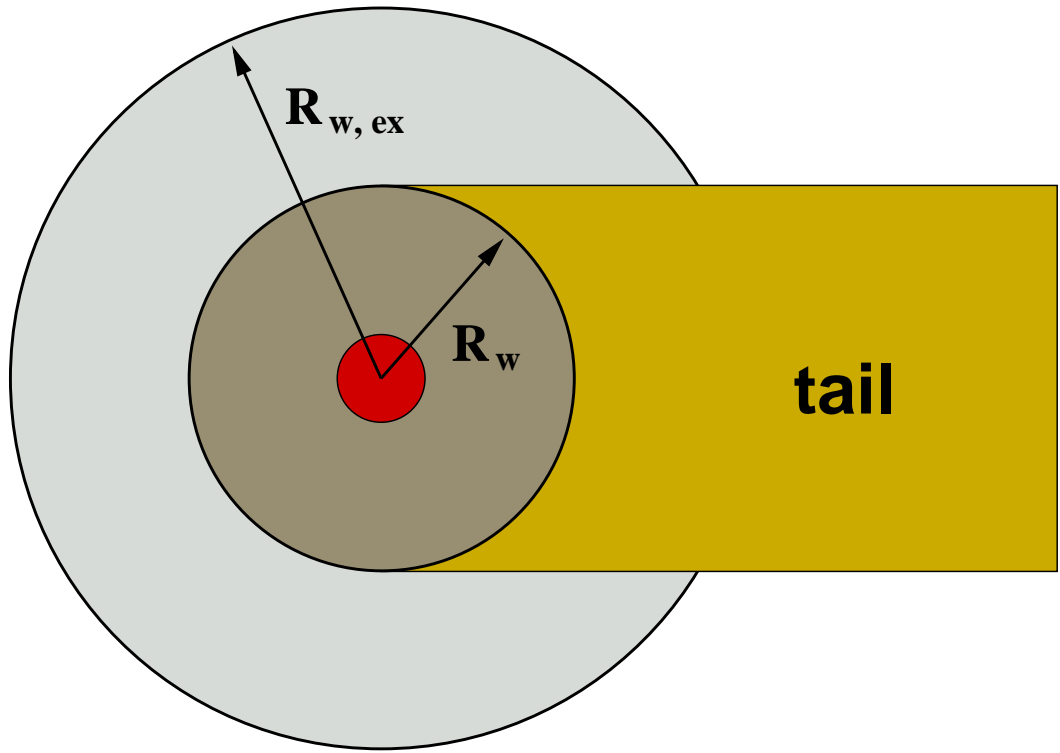}}}
\caption{A sketch of a stellar wind drawn out into a tail by the
radiation field of the central accretion disk, jet-base and jet.} 
\end{figure}
%%%%%%%%%%%%%%%%%%%%%%%%%%%%%%%%%%%%%%%%%

%%%%%%%%%%%%%%%to insert figure 7 %%%%%%%%
\begin{figure}
\centering\rotatebox{0}{\resizebox{14cm}{!}%
{\includegraphics{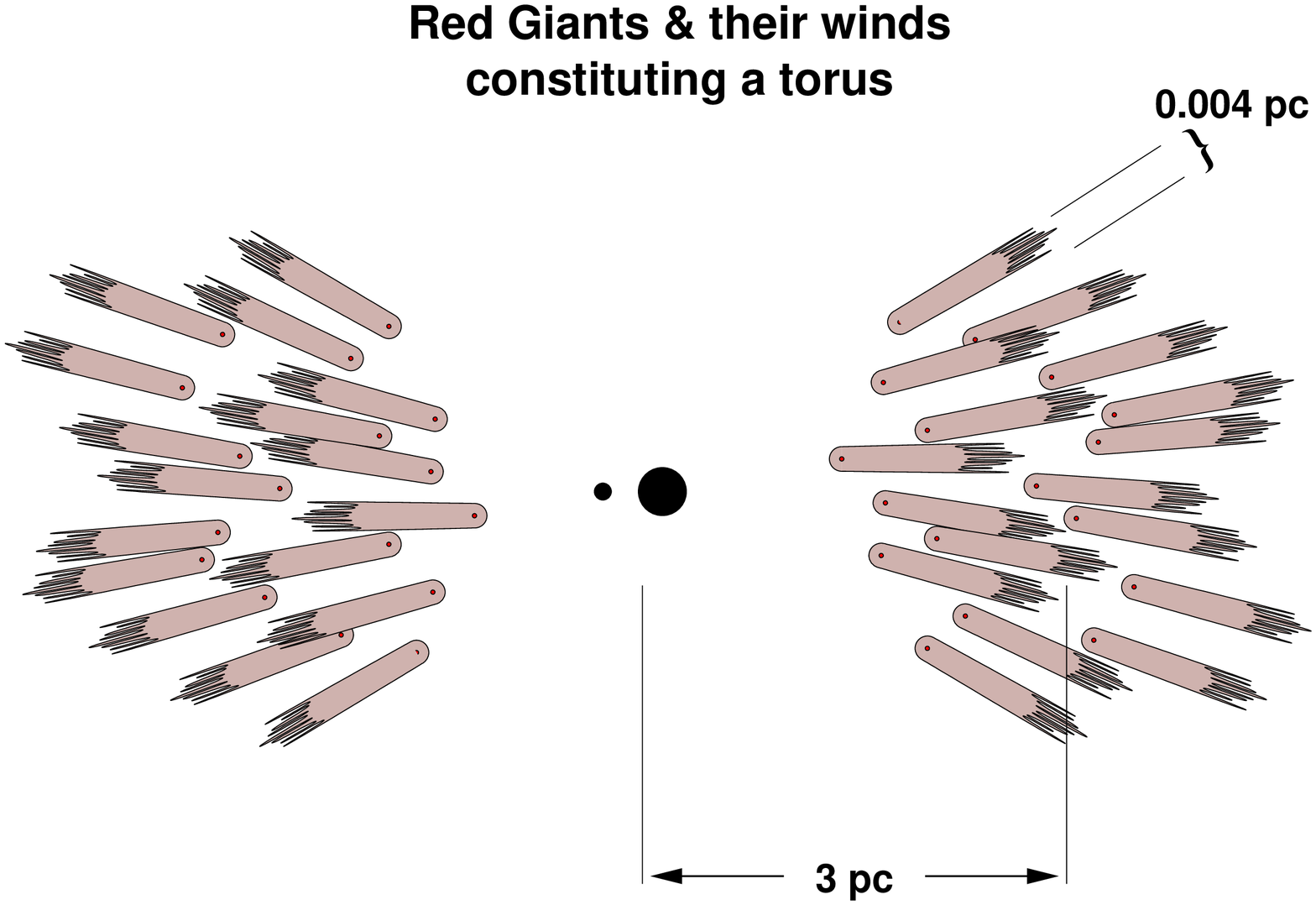}}}
\caption{A cross cut of the imagined ensemble of all the stars with wind
turned into tails constituting the torus.} 
\end{figure}
%%%%%%%%%%%%%%%%%%%%%%%%%%%%%%%%%%%%%%%%%

Such a torus like geometry for the stellar distribution is a natural
consequence of the torque exerted by a binary Black Hole system, see
figs. 4 and 5.  The central stellar distribution is carved out into a
broad belt, whose subpopulation of red and blue giant stars has their
winds turned into tails by radiation pressure from the central accretion
disk, see fig. 6.
\cite{BBH-2}.  These winds and their dust may constitute the ubiquituous
torus, see fig. 7.  Interestingly, this concept leads to a variety of
successful checks:  first, the number of stars necessary to remove
sufficient angular momentum from the central Black Hole binary is just
about equal to the number of stars implied by the necessity to be
geometrically and  optically thick (using the cross-section of all the
winds turned into tails).  Second, a single stellar wind-tail seen along
the tail can obscure the entire broad line region, thereby assuring a
rather short variability time scale, and also allowing for the partial
covering occasionally suggested by X-ray data.  Third, the column
inferred for wind-tails is of order $10^{23}$ to $10^{24}$ $cm^{-2}$, just
as required by the X-ray data as well,
\cite{BBH-2}.  On the other hand, the wind-tails exist only if there is
sufficient radiation from the inner accretion disk; should that disk have
turned to an ADAF, with little ultraviolet emission, such tails will not
exist, and so while the torus configuration in the stellar distribution
may exist, the obscuration may be quite different, certainly then with no
associated far infrared emission.

\subsection{Merger of two Black Holes: Spin-flip}

In recent work we have expanded this concept to include the final stages
of the merger, now between the two central Black Holes.  Fig. 8 shows our
basic concept of the spin-flip of two merging Black Holes, \cite{MCh02}.

%%%%%%%%%%%%%%%to insert figure 8 %%%%%%%%
\begin{figure}
\centering\rotatebox{-90}{\resizebox{11cm}{!}%
{\includegraphics{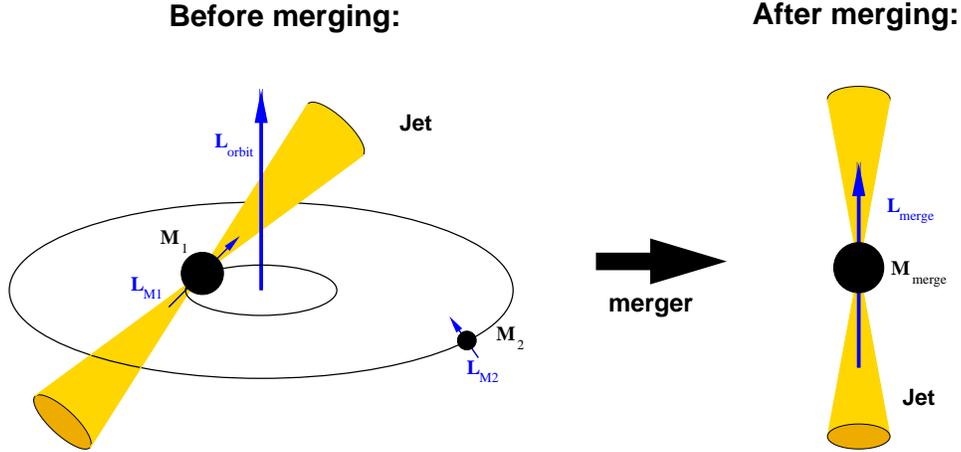}}}
\caption{The principle of the spin-flip of the more massive back
hole induced by the orbital angular momentum of the secondary black
hole.  This is a proposed solution to the riddle of the X-shaped radio
galaxies, \cite{Ro01}.} 
\end{figure}
%%%%%%%%%%%%%%%%%%%%%%%%%%%%%%%%%%%%%%%%%

When the two Black Holes get sufficiently close to induce gravitational
radiation, it becomes necessary to include the spin orientation of the
orbit of the two Black Holes, and their intrinsic spin.  Focussing on
the more massive Black Hole, and treating the second Black Hole in a test
particle limit, we have used a post-Newtonian approximation to derive the
properties of the gravitational waves, and the reaction of the two
orbiting Black Holes.  The interesting question is whether the angular
momentum transport by the gravitational wave emission will remove most of
the orbital angular momentum of the secondary Black Hole.  This would
thus lead to a final merged Black Hole whose total angular momentum has a
direction which is little modified from the pre-merger direction of the
spin of the primary Black Hole.  Also, one might expect that the spin as
a fraction of the maximum allowed is decreased relative to the pre-merger
state.

%%%%%%%%%%%%%%%to insert figure 9 %%%%%%%%
\begin{figure}
\begin{center}
\epsfig{file=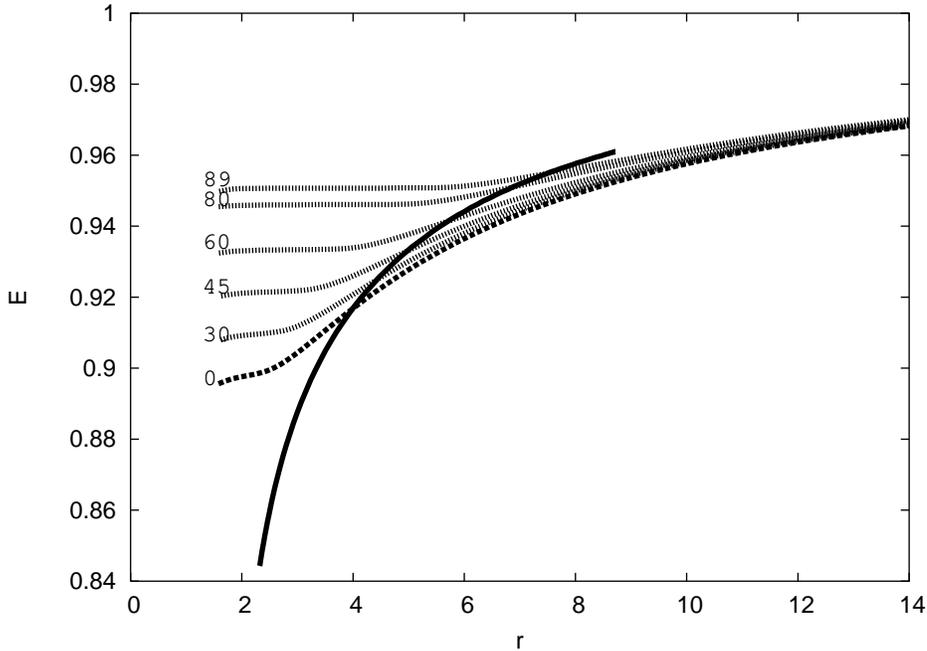}
\caption[]{The run of the energy for a test particle coming in along
different orbital planes; the dark line represents the innermost
stable orbit.  The innermost stable orbit as a function of
inclination has the shape of a symmetric gourd.  This figure shows that
upon passing the innermost stable orbit the energy loss gets very weak.} 
\end{center}
\end{figure}
%%%%%%%%%%%%%%%%%%%%%%%%%%%%%%%%%%%%%%%%%

Our calculations have shown that the opposite happens, see figs. 9 and
10:  the removal of orbital angular momentum has been demonstrated to
have a very strong dependence on the angle between orbital spin of the
secondary Black Hole and the intrinsic spin of the primary Black Hole. 
This dependence is so strong that the final Black Hole spin after the
merger is not only turned around, but also remains high relative to the
maximum spin allowed.  

This shows that a merger of a maximally rotating Black Hole with a
secondary Black Hole with an orbital spin axis which is in a very
different direction can lead to a spin flip, with the new spin again near
maximum. The key point is that the orbital
angular momentum of the secondary Black Hole is not completely removed by
gravitational waves, as one might have expected, see figs. 9 and 10.  Of
all proposed models, this, it turns out, is the best viable explanation
for the X-shaped radio galaxies, \cite{Ro01}.  

There is an important consequence, that any accretion disk again may go
very close to the horizon both before and after the merger, leading to a
very high temperature near the inner edge of the disk,
\cite{DoBi96,DoBi02}, especially should the formation of the jet lead to
an ADAF-like structure of that inner ring of the disk underneath the
budding jet \cite{DoBi96}.  This may be important for the hadronic
interactions in such a region,
\cite{CR-V,jd-2a}.

%%%%%%%%%%%%%%%to insert figure 10 %%%%%%%%
\begin{figure}
\begin{center}
\epsfig{file=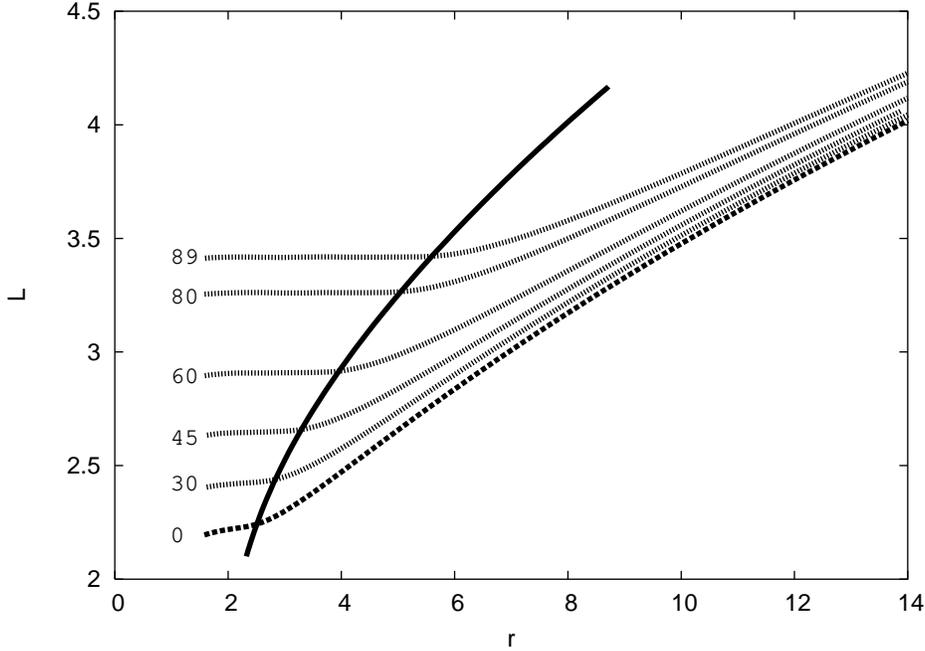}
\caption{The run of the angular momentum for a test particle coming in
along different orbital planes; the dark line represents the innermost
stable orbit.   This figure shows that upon passing the innermost stable
orbit the angular momentum loss gets approaches zero.} 
\end{center}
\end{figure}
%%%%%%%%%%%%%%%%%%%%%%%%%%%%%%%%%%%%%%%%%

Our recent calculations show that the spin dependence of the Black Hole
mergers needs to be taken into account because it severely modifies the
final gravitational wave emission.  This may have consequences for the
template used to look for the weak signal from Black Hole mergers.

\section{Summary}

Black Holes occur on many mass scales, spanning a factor of $10^9$.  We
have shown that the basic underlying physics of the jet-disk-Black Hole
system is very similar on all scales, from microquasars to the most
powerful quasars.  All appear to have a relativistic jet, a compact
accretion disk, and jet powers often equivalent if not larger than the
visible emission from the disk.  When galaxies merge, then by necessity
also two Black Holes form a binary, that in turn will merge after some
time.  The torque from this close Black Hole binary may produce a torus
like configuration among the inner stellar population; the winds of the
red and blue giant stars may constitute the torus, inferred from
X-ray, optical and mm-observations.  The final stages of the Black Hole
merger turns out to be strongly dependent on the relative spin direction
of the orbit and the more massive Black Hole.  This may have relevance for
any search for a Black Hole merger signal in gravitational waves.

\section{Acknowledgements}

P.L. Biermann would like to acknowledge the hospitality during his
prolonged stay at the University of Paris VII, in the spring 2002, offered
by his colleagues Norma Sanchez and Hector de Vega.  Work with PLB is
being supported by the through the AUGER theory and membership grant 05
CU1ERA/3 through DESY/BMBF; further support for the work with PLB comes
from the DFG, DAAD, Humboldt Foundation, KOSEF (Korea), and NSERC
(Australia).  Mihaela Chirvasa has been supported first through i) an
Erasmus grant from the University of Bucharest, and ii) the Max Planck
Institute for Radioastronomy, then by the University of Bucharest, and
now by the Max Planck Institute for Gravitational Physics.  S. Markoff
was supported by the Humboldt Foundation, and then also through the AUGER
grant.  S. Markoff and H. Falcke ackowledge their very intense
collaboration on questions disks and jets also with S. Corbel, R. Fender,
M. Nowak and F. Yuan.  P.L. Biermann and M. Chirvasa owe much to many
conversations with Gerd Sch{\"a}fer, University of Jena.  P.L.
Biermann and Ch. Zier would like to acknowledge long and intense
discussions over many years with H. Rottmann, whose Ph.D. thesis first
expounded upon the idea of a possible spin-flip of two merging black
holes.  P.L. Biermann and Ch. Zier also would like to acknowledge
intense and fruitful interactions with R. R. J. Antonucci.  P.L. Biermann
also would like to thank Marina Kaufman and Gustavo Romero for many
discussions of these topics, and then Albrecht Dress, Timo Kellmann,
and Athina Meli for help with the final editing of this ms.


\begin{thebibliography}{999}

\bibitem{AnMi85}  Spectropolarimetry and the nature of NGC 1068,
   Antonucci, R. R. J., \& Miller, J. S., \ApJ  {\bf  297}, 621-632 (1985)
%  Title:  Spectropolarimetry and the nature of NGC 1068

\bibitem{BiSt87}  Synchrotron emission from shock waves in active galactic
   nuclei, Biermann, P. L., \& Strittmatter, P. A., \ApJ  {\bf 322},
   643-649 (1987)
%  Title:  Synchrotron emission from shock waves in active galactic
%  nuclei

\bibitem{CR-V} Cosmic rays V. The nonthermal radioemission of the old nova
   GK Per - a signature of hadronic interactions?, Biermann, P.L., 
   Strom, R.G., \& Falcke, H., \AA {\bf 302}, 429 (1995),
   astro-ph/9508102
%  Title: Cosmic rays V. The nonthermal radioemission of the old nova
%  GK Per - a signature of hadronic interactions?

\bibitem{ESA-CERN01}  Astroparticles:  Cosmic Particle Physics, 
   Biermann, P.L., review and summary report, ESA-CERN Workshop on
   "Fundamental Physics in Space \& Related Topics", ESA report SP-469,
   Eds. M.C.E. Huber \& M. Jacob, 21 - 33 (2001)
%  Title:  Astroparticles:  Cosmic Particle Physics

\bibitem{Paris00}  Introduction to Cosmic Rays, Biermann, P.L.,
   \& Sigl, G., review for the Paris meeting, June 2000,
   ``Physics and Astrophysics of Ultra-High Energy Cosmic Rays", Eds. M.
   Lemoine, G. Sigl, Springer-Publ., p. 1 - 26 (2001)
%  Title:  Introduction to Cosmic Rays

\bibitem{Erice00}  High energy particles in Active Galactic Nuclei, 
   Biermann, P.L., review at the Erice school Nov. 2000,
   ``Astrophysical sources of high energy particles and radiation", Eds.
   M.M. Shapiro et al., Kluwer Acad. Publ., p. 115 - 133 (2001)
%  Title:  High energy particles in Active Galactic Nuclei

\bibitem{CR-APE02}  Cosmic Radiation, Biermann, P.L., \& Seo, E.-S.,
   review, Academic Press Encyclopedia, Third Edition, vol. 3,
   p. 823 - 835 (2002)
%  Title:  Cosmic Radiation

\bibitem{CKB89}  The nature of radio-quiet quasars, Chini, R., Kreysa, E.,
   \& Biermann, P. L., \AA {\bf  219}, 87 - 97 (1989)
%  Title:  The nature of radio-quiet quasars

\bibitem{CBKG89}  870 and 1300 micron observations of radio quasars,
   Chini, R., Biermann, P. L., Kreysa, E., \& Gemuend, H.-P., \AAL {bf
   221}, L3 - L6 (1989)
%  Title:  870 and 1300 micron observations of radio quasars

\bibitem{MCh02}  "Evolution of a test particle around a Kerr Black Hole
   under the emmision of gravitational waves", Chirvasa, M., M.Sc. thesis,
   University of Bucharest, February (2002)
%  Title:  Binary Black Holes

\bibitem{DMR00}  AU-Scale Synchrotron Jets and Superluminal Ejecta in GRS
   1915+105, Dhawan, V., Mirabel, I.~F., \& Rodr{\' i}guez, L.~F., \ApJ
   {\bf 543}, 373 - 385 (2000)
%  title = "{AU-Scale Synchrotron Jets and Superluminal Ejecta in GRS
%  1915+105}",

\bibitem{DoBi96}  The symbiotic system in quasars: black hole,
    accretion disk and jet, Donea, A.C., \& Biermann, P.L.,
    \AA, {\bf 316} 43 (1996), astro-ph/9602092
%   Title:  The symbiotic system in quasars: black hole,
%   accretion disk and jet

\bibitem{DoBi02}  The structure of accretion flow at the base of jets in
   AGN, Donea, A.-C., \& Biermann, P.L., , {\it Publ. of the Astron. Soc.
   of Australia (PASA)} {\bf 19}, 125 - 128 (2002)
%  Title:  The structure of accretion flow at the base of jets in AGN

\bibitem{DSB00} Viscosity in self-gravitationg disks, Duschl, W.J., 
   Strittmatter, P.A., \& Biermann, P.L., \AA {\bf 357}, 1123 - 1132 
   (2000), astro-ph/0004076
%  Title:  Viscosity in self-gravitationg disks

\bibitem{EWNB98}  Black hole energy release to the gaseous universe,
   En{\ss}lin, T.A., Wang, Y., Nath, B.B., \& Biermann, P.L., \AAL   {\bf
   333}, L47 - L50 (1998), astro-ph/9803105
%  Title:   Black hole energy release to the gaseous universe

\bibitem{FTA97}  The Centers of Early-Type Galaxies with HST. IV. Central
   Parameter Relations, Faber, S. M., Tremaine, S., Ajhar, E. A.
   \etal, \AJ   {\bf 114}, 1771 (1997)
%  Title:  The Centers of Early-Type Galaxies with HST. IV. Central
%  Parameter Relations

\bibitem{jd-1} The jet-disk symbiosis I. Radio to X-ray emission models
   for quasars, Falcke, H., \& Biermann, P.L.,  \AA {\bf 293}, 665,
   1995, astro-ph/9411096
%  Title:   The jet-disk symbiosis I. Radio to X-ray emission models for
%  quasars

\bibitem{jd-2} The jet-disk symbiosis II. Interpreting the radio/UV
   correlations in quasars, Falcke, H., Malkan, M.A., \& Biermann, P.L.,
   \AA {\bf 298}, 375 (1995), astro-ph/9411100
%  Title:  The jet-disk symbiosis II. Interpreting the radio/UV
%  correlations in quasars

\bibitem{jd-2a} Unified schemes for active galaxies: a clue from the
   missing Fanaroff-Riley type I quasar population, Falcke, H.,
   Krishna, G., \& Biermann, P.L., \AA {\bf 298}, 395 (1995),
   astro-ph/9411106
%  Ttle:  Unified schemes for active galaxies: a clue from the missing
%    Fanaroff-Riley type I quasar population

\bibitem{jd-3}  The jet/disk symbiosis III. What the radio cores in
   GRS1915+105, NGC4258, M81 and Sgr A$^{\star}$ tell us about accreting
   black holes, Falcke, H., \& Biermann, P.L., \AA {\bf 342}, 49 - 56
   (1999, astro-ph/9810226
%  Title:  The jet/disk symbiosis III. What the radio cores in
%  GRS1915+105, NGC4258, M81 and Sgr A$^{\star}$ tell us about accreting
%  black holes

\bibitem{FaMa00} The jet model for Sgr A*: Radio and X-ray spectrum, 
   Falcke, H., \& Markoff, S., \AA {\bf 362},  113 - 118 (2000)
%  Title:  The jet model for Sgr A*: Radio and X-ray spectrum

\bibitem{FeMe00}  A Fundamental Relation between Supermassive Black Holes
   and Their Host Galaxies, Ferrarese, L., \& Merritt, D., \ApJL  {\bf
   539}, L9 - L12 (2000) 
%  title = "{A Fundamental Relation between Supermassive Black Holes
%  and Their Host Galaxies}",

\bibitem{KBRM02}  Precessing microblazars and unidentified gamma-ray
   sources, Kaufman Bernad{\' o}, M.~M., Romero, G.~E., \& Mirabel,
   I.~F.,  \AAL {\bf 385}, L10 - L13 (2002)
%  title = "{Precessing microblazars and unidentified gamma-ray
%  sources}",
   
\bibitem{KFM02}  Population X: Are the super-Eddington X-ray sources
   beamed jets in microblazars or intermediate mass black holes?,
   K{\"o}rding, E., Falcke, H., \& Markoff, S., \AAL {\bf 382}, L13 - L16
   (2002)
%  Title:  Population X: Are the super-Eddington X-ray sources beamed
%  jets in microblazars or intermediate mass black holes?

\bibitem{KoRi95}  Inward Bound---The Search For Supermassive Black Holes
   In Galactic Nuclei, Kormendy, J., \& Richstone, D., \ARAA {\bf  33},
   581 (1995)
%  Title:  Inward Bound---The Search For Supermassive Black Holes In
%  Galactic Nuclei

\bibitem{LaEl82}  Obscuration and the various kinds of Seyfert
   galaxies, Lawrence, A., \& Elvis, M., \ApJ  {\bf  256}, 410 - 426
   (1982)
%  Title:  Obscuration and the various kinds of Seyfert
%  galaxies

\bibitem{LeMa00}  High-Energy Neutrino Astrophysics, Learned, J. G., \& 
   Mannheim, K., \ARNPS  {\bf 50}, 679 - 749 (2000)
%  Title:  High-Energy Neutrino Astrophysics

\bibitem{LRB98}  Magnetically supported tori in active galactic nuclei,
   Lovelace, R.V.E., Romanova, M., \& Biermann, P.L., \AA {\bf 338}, 856 
   (1998)
%  Title:  Magnetically supported tori in active galactic nuclei

\bibitem{MFF01}  A jet model for the broadband spectrum of XTE J1118+480.
   Synchrotron emission from radio to X-rays in the Low/Hard spectral
   state,  Markoff, S., Falcke, H., \& Fender, R., \AA {\bf 372}, L25 -
   L28 (2001)
%  Title:  A jet model for the broadband spectrum of XTE J1118+480.
%  Synchrotron emission from radio to X-rays in the Low/Hard spectral
%  state

\bibitem{MFYB01}  The Nature of the 10 kilosecond X-ray flare in Sgr A*,
   Markoff, S., Falcke, H., Yuan, F., \& Biermann, P.L., \AA {\bf 379},
   L13 - L16 (2001)
%  Title:   The Nature of the 10 kilosecond X-ray flare in Sgr A*

\bibitem{MNCFF02}  Exploring the Role of Jets in the Radio/X-ray
   Correlations of GX339-4, Markoff, S., Nowak, M., Corbel, S., Fender,
   R., \& Falcke, H., \AA {\bf  } (in press) (2002); astro-ph/0210439 
%  Title:  Exploring the Role of Jets in the Radio/X-ray Correlations
%  of GX339-4

\bibitem{YMFB02}  NGC 4258: A jet-dominated low-luminosity AGN?, Yuan,
   F., Markoff, S., Falcke, H., \& Biermann, P.L., \AA {\bf 391}, 139 -
   148  (2002)
%  Title:  NGC 4258: A jet-dominated low-luminosity AGN?

\bibitem{MeFa01}  The Supermassive Black Hole at the Galactic Center, 
   Melia, F., \& Falcke, H., \ARAA {\bf 39}, 309 - 352 (2001)
%  Title:   The Supermassive Black Hole at the Galactic Center

\bibitem{MeEk02}  Tracing Black Hole Mergers Through Radio Lobe
   Morphology, Merritt, D., \& Ekers, R.~D., {\it Science}  {\bf  297},
   1310 - 1313 (2002)
%  title = "{Tracing Black Hole Mergers Through Radio Lobe Morphology}",

\bibitem{Me02}  Rotational Brownian Motion of a Massive Binary, Merritt,
   D., \ApJ  {\bf  568}, 998 - 1003  (2002)
%  title = "{Rotational Brownian Motion of a Massive Binary}",

\bibitem{MeFe01}  Black hole demographics from the $M_{BH}-{\sigma} $
   relation, Merritt, D., \& Ferrarese, L., \MNRAS  {\bf  320},
   L30 - L34 (2001)
%  title = "{Black hole demographics from the M_{&bull;}-{\sigma}
%  relation}",

\bibitem{MMRB02}  Galaxy cores as relics of black hole mergers, 
   Milosavljevi{\'c}, M.~, Merritt, D., Rest, A., \& van den Bosch,
   F.~C., \MNRAS  {\bf 331}, L51 - L55 (2002)
%  title = "{Galaxy cores as relics of black hole mergers}",

\bibitem{MDMRG01}  A high-velocity black hole on a Galactic-halo orbit in
   the solar neighbourhood, Mirabel, I.~F., Dhawan, V., Mignani,
   R.~P., Rodrigues, I., \& Guglielmetti, F., \Nature  {\bf  413}, 139
   - 141 (2001)
%  title = "{A high-velocity black hole on a Galactic-halo orbit in
%  the solar neighbourhood}",
 
\bibitem{MiRo00}  Microquasars, Mirabel, I.~F., \& Rodr{\' i}guez, L.~F.,
   {\it Nucl. Phys. B Proc. Suppl.}  {\bf 80}, 143 - 151 (2000)
%   title = "{Microquasars}",

\bibitem{MiRo99}  Sources of Relativistic Jets in the Galaxy, Mirabel,
   I.~F., \& Rodr{\' i}guez, L.~F., \ARAA {\bf 37}, 409 - 443 (1999)
%  title = "{Sources of Relativistic Jets in the Galaxy}",

\bibitem{MDC98}  Accretion instabilities and jet formation in GRS
   1915+105, Mirabel, I.~F., Dhawan, V., Chaty, S., \etal, \AAL {\bf 
   330}, L9 - L12 (1998)
%  title = "{Accretion instabilities and jet formation in GRS
%  1915+105}",

\bibitem{MiRo98}  Microquasars in our Galaxy, Mirabel, I.~F., 
   \& Rodriguez, L.~F., \Nature  {\bf  392}, 673 - 676 (1998)
%  title = "{Microquasars in our Galaxy.}",
 
\bibitem{Mu82}  The X-ray spectrum and time variability of narrow emission
   line galaxies, Mushotzky, R. F., \ApJ  {\bf  256}, 92 - 102 (1982)
%  Title:  The X-ray spectrum and time variability of narrow emission
%  line galaxies

\bibitem{PeLe96}  The Fluence Distribution of Gamma-Ray Bursts, Petrosian,
   V., \& Lee, T. T., \ApJL  {\bf 467}, L29 (1996)
%  Title:  The Fluence Distribution of Gamma-Ray Bursts

\bibitem{PiKr92}  Infrared spectra of obscuring dust tori around active
   galactic nuclei. I - Calculational method and basic trends, Pier, E.
   A., \& Krolik, J. H., \ApJ  {\bf  401}, 99 - 109 (1992) 
%  Title:  Infrared spectra of obscuring dust tori around active galactic
%  nuclei. I - Calculational method and basic trends

\bibitem{PiKr93}  Infrared Spectra of Obscuring Dust Tori around Active
   Galactic Nuclei. II. Comparison with Observations, Pier, E. A., \& 
   Krolik, J. H., \ApJ  {\bf  418}, 673 (1993)
%  Title:  Infrared Spectra of Obscuring Dust Tori around Active Galactic
%  Nuclei. II. Comparison with Observations

\bibitem{GRB-1}  A jet-disk symbiosis model for Gamma Ray Bursts:  SS 433
   the next?, Pugliese, G., Falcke, H., \& Biermann, P.L., \AAL {\bf
   344}, L37 - L40 (1999), astroph/9903036
%  Title:  A jet-disk symbiosis model for Gamma Ray Bursts:  SS 433 the
%  next?

\bibitem{GRB-2}  The jet-disk model for GRBs:  cosmic ray and neutrino
   backgrounds, Pugliese, G., Falcke, H., Wang, Y., \& Biermann, P.L.,
   \AA {\bf 358}, 409 - 416 (2000), astro-ph/0003025
%  Title:   The jet-disk model for GRBs:  cosmic ray and neutrino
%  backgrounds

\bibitem{RKBM02}  Recurrent microblazar activity in Cygnus X-1?, Romero,
   G.~E., Kaufman Bernad{\' o}, M.~M., \& Mirabel, I.~F., \AAL {\bf 393},
   L61  -L64 (2002)
%  Title:  Recurrent microblazar activity in Cygnus X-1?
   
\bibitem{Ro01}  Ph.D. Thesis, Rottmann, H., Univ. of Bonn, (2001)

\bibitem{SaMi96}  Luminous Infrared Galaxies, Sanders, D. B., \& Mirabel,
   I. F., \ARAA {\bf 34}, 749 (1996)
%  Title:  Luminous Infrared Galaxies

\bibitem{SPN89}  Continuum energy distribution of quasars - Shapes and
   origins,  Sanders, D. B., Phinney, E. S., Neugebauer, G., \etal, 
\ApJ 
   {\bf  347}, 29-51 (1989) 
%  Title:  Continuum energy distribution of quasars - Shapes and
%  origins

\bibitem{WBW00}  Black hole to bulge mass correlation in Active Galactic
   Nuclei:  A test for a simplified formation scheme, Wang, Y., Biermann,
   P.L., \& Wandel, A., \AA {\bf  361}, 550 - 554 (2000),
   astro-ph/0008105
%  Title:  Black hole to bulge mass correlation in Active Galactic
%  Nuclei:  A test for a simplified formation scheme

\bibitem{WaBi98}  A possible mechanism for the mass ratio limitation in
   early type galaxies, Wang, Y., \& Biermann, P.L., \AA {\bf 334}, 87 -
   95 (1998), astroph/9801316
%  Title:  A possible mechanism for the mass ratio limitation in early
%  type galaxies

\bibitem{WaBi00}  Merger-driven galaxy evolution, faint IR source counts,
   and the background, Wang, Y., \&  Biermann, P.L., \AA {\bf 356}, 808 -
    (2000), astro-ph/0003005
%  Title:  Merger-driven galaxy evolution, faint IR source counts, and
%  the background

\bibitem{YMF02}  A Jet-ADAF model for Sgr A*, Yuan, F.,  Markoff, S.,
   Falcke, H., \AA {\bf 383}, 854 - 863 (2002)
%  Title:  A Jet-ADAF model for Sgr A*

\bibitem{Zi00}  Ph.D. thesis, Zier. Ch., Univ. Bonn, 2000

\bibitem{BBH-1}  Binary Black Holes and Tori in AGN I. Ejection of stars
   and merging of the binary, Zier, Ch., \& Biermann, P.L., \AA  {\bf
   377}, 23 - 43 (2001)
%  Title:  Binary Black Holes and Tori in AGN I. Ejection of stars
%  and merging of the binary

\bibitem{BBH-2}   Binary Black Holes and tori in AGN II. Can stellar winds
   constitute a dusty torus?, Zier, Ch., \& Biermann, P.L., \AA  {\bf  }
  (in press) (2002)
%  Binary Black Holes and tori in AGN II. Can stellar winds
%  constitute a dusty torus?

\end{thebibliography}
\end{document}